\documentclass[aps,prd,12point,twocolumn,nofootinbib,showpacs,superscriptaddress]{revtex4-1}
\def\theequation{\arabic{section}.\arabic{equation}}
\usepackage{psfrag}
\usepackage{subfigure}
\usepackage{color}
\usepackage{mathrsfs}
\usepackage{graphicx}
\usepackage{amssymb, bm}
\usepackage{amsmath, amsthm}
\usepackage{epstopdf}
\usepackage{hyperref}
\usepackage{enumerate}
\usepackage{longtable}

\newcommand{\be}{\begin{equation}}
\newcommand{\ee}{\end{equation}}

\begin{document}
\def\theequation{\arabic{section}.\arabic{equation}} 

\title{Scalar field as a null dust}

\author{Valerio Faraoni}
\email[]{vfaraoni@ubishops.ca}
\affiliation{Department of Physics and Astronomy and {\em STAR} Research 
Cluster, Bishop's University, 2600 College Street, Sherbrooke, Qu\'ebec, 
Canada J1M~1Z7 }

\author{Jeremy C\^ot\'e}
\email[]{jcote16@ubishops.ca}
\affiliation{Department of Physics and Astronomy, Bishop's University, 
2600 College Street, Sherbrooke, Qu\'ebec, 
Canada J1M~1Z7}



\begin{abstract}

We show that a canonical, minimally coupled scalar field which is non-self 
interacting and massless is equivalent to a null dust fluid (whether it is 
a test or a gravitating field), in a spacetime region in which its 
gradient is null. Under similar conditions, the gravitating and 
nonminimally coupled Brans-Dicke-like scalar of scalar-tensor gravity, 
instead, cannot be represented as a null dust unless its gradient is also 
a Killing vector field.

\end{abstract}


\maketitle

\section{Introduction}
\label{sec:1}
\setcounter{equation}{0}

Scalar fields give rise to the simplest field theory and  are 
ubiquitous in cosmology, particle physics, and theories and models of 
classical and quantum gravity \cite{Waldbook, Carroll, Bojowald, Linde, 
AmendolaTsujikawabook, reviews}. A 
minimally coupled scalar field $\phi$ 
with canonical kinetic energy is 
described by the action 
\be 
S_{(\phi)}=\int d^4 x \sqrt{-g} \left[ 
-\frac{1}{2} \nabla^c \phi \nabla_c \phi -V(\phi) \right]  \,,
\ee 
where $g$ is the determinant of the spacetime metric $g_{ab}$,
$\nabla_a$ denotes the covariant derivative operator of $g_{ab}$, and 
$V(\phi)$ is the scalar field potential (we use 
units in which the speed of light and Newton's constant are unity and we 
follow the notation of Ref.~\cite{Waldbook}).  The scalar 
field stress-energy tensor is
\be T_{ab}^{(\phi)} = -\frac{2}{\sqrt{-g}} \, \frac{ \delta 
S_{(\phi)}}{\delta g^{ab}} =\nabla_a \phi \nabla_b \phi 
-\frac{1}{2} \, g_{ab} \nabla^c \phi \nabla_c \phi -V(\phi) g_{ab}  
\label{set}
\ee 
and $\phi$ satisfies the Klein-Gordon equation obtained from the covariant 
conservation $\nabla^b T_{ab}^{(\phi)} =0$,
\be
\Box \phi -\frac{dV}{d\phi} =0 \,, \label{KG}
\ee
where $\Box \equiv g^{ab} \nabla_a \nabla_b $ is the curved space 
d'Alembertian. 
The question of whether this scalar field can be represented as an 
effective fluid has been posed, and answered, long ago 
\cite{Ellis71, Madsen, myphifluid}.  If 
the 
gradient $\nabla^c \phi$ is timelike in a spacetime region, the scalar 
field stress-energy tensor~(\ref{set}) is equivalent to a perfect fluid 
stress-energy tensor \cite{Ellis71, Madsen, myphifluid} with fluid  
four-velocity 
\be 
u_a=\frac{\nabla_a \phi}{ \sqrt{-\nabla^c \phi \nabla_c \phi} } \,. 
\ee
Spacelike scalar field gradients have also been considered 
\cite{myphifluid, Semiz}, 
although they are not physically very relevant. Here we focus on the 
remaining case, apparently not covered in the literature, in which 
$\nabla^c \phi$ is null on a  spacetime region, 
$\nabla^c\phi \nabla_c\phi=0$. A {\em dust} corresponds to a  fluid with 
timelike four-velocity $u^a$, energy density $\rho$, and zero pressure 
described by the energy-momentum tensor  $T_{ab}^{(dust)}=\rho \, u_a u_b 
$. Because there is no pressure gradient, the fluid elements of the dust 
follow timelike geodesics, as can be deduced from 
covariant conservation \cite{Waldbook}.  
A {\em null dust} \cite{KSMcH, KucharBicak} corresponds to the limit in 
which the four-velocity becomes null, and is described by 
\be
T_{ab}^{(nd)}= \rho \, k_a k_b \,, \;\;\;\;\;\;\;\;\; k_c k^c=0 \,, 
\label{dust-set}
\ee
where $\rho>0$ and the trace $T \equiv {T^c}_c$ vanishes. The 
covariant 
conservation of $T_{ab}^{(nd)}$ implies that $k^c $ is also geodesic 
\cite{KSMcH, KucharBicak}. The null dust is interpreted as a 
coherent zero rest mass field 
propagating at the speed of light in the null direction $k^a$. Naturally, 
a null dust can be realized by propagating electromagnetic 
\cite{Robinson} or gravitational \cite{Griffiths, KSMcH} waves. Here we 
consider the analogous 
problem for a massless scalar field.  The null 
dust plays a non-negligible role in the literature on classical and 
quantum gravity, especially in the study of 
Vaidya \cite{Vaidya}, $pp$-wave \cite{KSMcH, Krasinski, Griffiths},  
Robinson-Trautman 
\cite{RobinsonTrautman}, and twisting \cite{twisting}  
solutions of the Einstein or 
Einstein-Maxwell equations, classical and quantum  gravitational  collapse,  
horizon formation, mass inflation \cite{exactcollapse}, black 
hole evaporation \cite{evaporation}, and 
canonical Lagrangians and Hamiltonians \cite{KucharBicak, Horvath, Myers}. 
More recently, null dust has 
been studied in relation with the fluid-gravity correspondence and 
holography \cite{various, Minwalla, Myers}. The collision of special  
scalar field-null dust solutions was studied long ago in \cite{ndsf} and  
scalar-Vaidya solutions are of interest in the AdS/CFT correspondence 
\cite{Minwalla, Myers} .   
Since the null vector $k^a$ can be rescaled by a positive function without 
changing its causal character, it is possible to find a representation of 
the stress-energy tensor~(\ref{dust-set}) in which $\rho \, \dot{=} \, 1$ 
and 
$T_{ab}^{(nd)} \, \dot{=} \, k_a k_b$ (a dot over an equal sign 
denotes the fact that the equality is only valid in that representation). 
However, in general, in this representation, the null 
geodesics tangent to $k^a$ are not affinely parametrized unless $k^a$ is 
divergence-free \cite{KucharBicak}.   

A natural question arises: is a scalar field $\phi$ with null gradient 
in a region of spacetime equivalent to an effective null dust? Special 
solutions with this property are known \cite{ndsf, Griffiths}. As 
expected, the answer is affirmative but only if $\phi$ is massless and 
there is no potential $V(\phi)$. Moreover, being generated by a scalar, 
this 
effective null dust is irrotational, as expected. A second question also 
arises naturally. 
Given that scalar fields play an important role in theories of gravity 
alternative to Einstein's general relativity (GR) searched for in tests of 
gravity \cite{tests}, 
the simplest 
alternative being Brans-Dicke theory \cite{BD} and its scalar-tensor 
generalizations \cite{ST} which contain a  gravitational scalar field 
$\Phi$ (approximately 
equivalent to the inverse of the effective gravitational 
coupling strength), is it possible that this scalar is equivalent to an 
effective null 
dust if $\nabla^c\Phi \nabla_c \Phi=0$? The scalar-tensor 
action is \cite{BD, ST} 
\begin{eqnarray}
S_{(ST)} &=& \frac{1}{16\pi} \int d^4x \sqrt{-g} \left[ \Phi R 
-\frac{\omega(\Phi )}{\Phi} 
\, \nabla^c\Phi \nabla_c\Phi -V(\Phi) \right] \nonumber\\
&&\nonumber\\
&\, & +S_{(m)} \,, \label{STaction}
\end{eqnarray}
where  the function $\omega(\Phi)$ (which was a  
constant parameter in the original Brans-Dicke theory 
\cite{BD}) is the ``Brans-Dicke coupling'', $V(\Phi)$ is a 
scalar field potential (absent in the original Brans-Dicke theory), 
and $S_{(m)}=\int d^4x \sqrt{-g} \, {\cal L}_{(m)} $ describes the 
matter sector of the theory. The scalar-tensor field equations 
obtained from the variation of this action take the form of effective 
Einstein equations,
\begin{eqnarray}
&&R_{ab} - \frac{1}{2}\, g_{ab} R = \frac{8\pi}{\Phi} \,  T_{ab}^{(m)} 
\nonumber\\
&&\nonumber\\
&\, & + \frac{\omega}{\phi^2} \left( \nabla_a \Phi 
\nabla_b \Phi -\frac{1}{2} \, g_{ab} 
\nabla_c \Phi \nabla^c \Phi \right) \nonumber\\
&&\nonumber\\
&\, &  +\frac{1}{\Phi} \left( \nabla_a \nabla_b \Phi 
- g_{ab} \Box \Phi \right) 
-\frac{V}{2\Phi}\, 
g_{ab} \,, \nonumber\\
&& \label{BDfe1} \\
&& \Box \Phi = \frac{1}{2\omega+3}   
\left( 
\frac{8\pi T^{(m)} }{\Phi}   + \Phi \, \frac{d V}{d\Phi} 
-2V -\frac{d\omega}{d\Phi} \nabla^c \Phi \nabla_c \Phi \right) \,, 
\nonumber\\
&& \label{BDfe2}
\end{eqnarray}
where $ T^{(m)} \equiv g^{ab}T_{ab}^{(m)} $ is the trace of the matter 
stress-energy tensor $T_{ab}^{(m)} $.  It is well known that, if the 
gradient $\nabla^c\Phi$ is timelike, the Brans-Dicke-like scalar $\Phi$ 
acts as an effective fluid source which, however, is not a perfect fluid 
as in the minimally coupled case, but is instead an (effective) imperfect 
fluid with a heat 
flux density, described by an effective stress-energy tensor of the form 
\cite{Madsen, Pimentel89, Kolassisetal88, Pimenteletal6, FaraoniCote}
\be
T_{ab}^{(\Phi)}= \left( P_{(\Phi)}+\rho_{(\Phi)} \right) u_a u_b + 
P_{(\Phi )} 
g_{ab} + q_a^{(\Phi)} u_b + q_b^{(\Phi)} u_a \,. \label{Phiset}
\ee
When $\nabla^c \Phi \nabla_c \Phi<0$, the effective fluid 
four-velocity is  
\be
u_a = \frac{ \nabla_a \Phi}{ \sqrt{ -\nabla^c \Phi \nabla_c \Phi}} \,, 
\;\;\;\;\;\; u_c u^c=-1 \,,
\ee
and 
\be
q_c^{(\Phi)} q^c_{(\Phi)}  >  0 \,, \;\;\;\;\;\; q_c^{(\Phi)} u^c=0 \,.
\ee
Explicit expressions of the effective fluid quantities $ \rho_{\Phi} 
, P_{\Phi}$, and $q_a^{(\Phi)} $ are given in Refs.~\cite{Pimentel89, 
FaraoniCote}.

An important conceptual step has been taken in moving the question  from 
the minimally coupled scalar $\phi$ to the Brans-Dicke-like field $\Phi$: 
the 
former can be a test field or  a matter source of the Einstein equations, 
while the latter always contributes to sourcing the metric in the 
scalar-tensor field equations~(\ref{BDfe1}). This distinction will have to 
be kept in mind in the following sections.

It turns out that, contrary to its minimally coupled counterpart $\phi$, a 
Brans-Dicke-like field $\Phi$ cannot be regarded as a null dust, which is 
a 
perfect fluid. In fact, in the stress-energy $T_{ab}^{(\Phi)}$ given by 
Eq.~(\ref{Phiset}), the terms $\nabla_a \nabla_b \Phi, \Box\Phi$ linear in 
the second covariant derivatives of $\Phi$ always introduce an (effective) 
imperfect fluid component, {\em i.e.}, a heat flux. By contrast, the 
canonical terms $\nabla_a \Phi \nabla_b \Phi , \nabla^c \Phi \nabla_c \Phi 
$ quadratic in the first order covariant derivatives correspond to 
(effective) perfect fluid terms.

The effective imperfect fluid description of scalar-tensor gravity has 
been applied recently to elucidate anomalies in the limit to GR of 
electrovacuum Brans-Dicke theory \cite{ourBDlimit}. However, it is not 
possible to do so for the corresponding scalar-tensor solutions describing 
null fields because a null dust description of scalar-tensor gravity is 
missing in this case.

\section{Massless canonical scalar field as an effective null dust in GR} 
\label{sec:2}
\setcounter{equation}{0}

In this section we restrict to Einstein's theory. In GR, a canonical 
scalar field (whether it is a test field or a gravitating 
one) has stress-energy tensor~(\ref{set}) and satisfies the Klein-Gordon 
equation~(\ref{KG}). We assume that $ \nabla^c
\phi\nabla_c \phi =0$ in a certain spacetime region, to which we will 
implicitely limit ourselves in the rest of this work. Then, the 
requirement that $T_{ab}^{(\phi)}$ assumes the null dust form  
$T_{ab}=\rho \, k_a k_b$ with $k^c$ null, necessarily implies $ 
T^{(\phi)}=0$ and $V(\phi)=0$, as expected. Then, the Klein-Gordon 
equation reduces to $\Box \phi=0$ and~(\ref{set}) reduces to 
$T_{ab}^{(\phi)}= \nabla_a \phi \nabla_b \phi$. {\em A priori}, there are 
two possibilities to identify this stress-energy tensor with that of a 
null dust. One could choose the representation in which 
\be
T_{ab}^{(\phi)} = \nabla_a \phi \nabla_b \phi \, \dot{=} \, k_a k_b \,, 
\;\;\;\;\; k_a \, \dot{=} \, \nabla_a \phi \,, \;\;\;\; \rho \, \dot{=} 
\, 1 \,.
\ee
For a general null dust, this representation excludes the affine 
parametrization for the null geodesics tangent to $k^a$. However, here 
these two representations become compatible because of the irrotationality 
of this scalar field-dust. 
In fact, the covariant conservation equation $\nabla^b T_{ab}^{(\phi)}=0$ 
yields
\be
k^b \nabla_b k_c = -\left( \nabla^b k_b \right) k_c \,,\label{Delta}
\ee
but the divergence of $k^c $ vanishes because $k_c$ is a pure gradient. 
Indeed, $ \nabla^a k_a = \Box \phi=0$ by 
virtue of the Klein-Gordon equation and Eq.~(\ref{Delta}) reduces to the 
affinely parametrized geodesic equation $k^b \nabla_b k_c =0$. For a 
scalar field-null dust, therefore, the representation with unit energy 
density $\rho=1$ coincides with the affine parametrization along the 
null geodesics tangent to $k^a$, due to irrotationality.

The second {\em a priori} possibility consists of keeping $\rho$ general 
and choosing
\be
T_{ab}^{(\phi)} = \nabla_a\phi \nabla_b\phi =\rho \, k_a k_b  \,, 
\;\;\;\;\;   k_a = \frac{\nabla_a \phi}{ \sqrt{\rho}} \,.
\ee
Covariant conservation then gives
\be
\nabla^b T_{ab}^{(\phi)} =\left( k^b \nabla_b \rho + \rho \nabla^b k_b 
\right) k_a +\rho k^b \nabla_b k_a =0 \,.\label{noh}
\ee
Using the Klein-Gordon equation $\Box\phi=0$, we now have 
\be
\nabla^a k_a = - 
\frac{ \nabla^a \rho \nabla_a \phi}{ 2\rho^{3/2}} 
\ee
so that Eq.~(\ref{noh}) becomes
\be
k^b \nabla_b k_c =\frac{1}{\rho} \left(  - k^b \nabla_b \rho +\frac{ k^b 
\nabla_b \rho }{2\sqrt{\rho}} \right) k_c \,.
\ee
Also in this representation it could seem that affine parametrization is 
incompatible with the choice $\rho \equiv 1$ but, again, this is not the 
case. Let us adopt the representation $\rho \, \dot{=} \, 1$; then we have
$ k^c k_c=0 $, 
\be
k^a\nabla_b k_a =0 
\ee
(which follows from taking the covariant derivative of 
the normalization $k^c k_c=0$), and
\be
\nabla^c k_c=0 \,.
\ee
At this point is is easy to derive also a wave equation for the null 
vector field $k^c$, 
\be
\Box k_a -R_{ab} k^b =0 \,,\label{eq:wv}
\ee
in which $k^c=\nabla^c \phi $ couples explicitly to the curvature (to the 
Ricci tensor  $R_{ab}$)  even 
though $\phi$ does not couple (to the Ricci scalar $R$) in the 
Klein-Gordon equation. To 
obtain Eq.~(\ref{eq:wv}), first note that
\be
\Box k_a = \nabla^b \nabla_b k_a=  \nabla^b \left( \nabla_b \nabla_a \phi 
\right) = \nabla^b \left( \nabla_a \nabla_b \phi \right) =
\nabla^b\nabla_a k_b 
\ee
and that the identity \cite{Waldbook}
\be
\Big[ \nabla_a, \nabla_b \Big] k_c = {R_{abc}}^d k_d=R_{abcd} k^d 
\ee
yields 
\be
\nabla_b\nabla_a  k_c = \nabla_a\nabla_b k_c - R_{abcd} k^d\,,
\ee
so that 
\be
\nabla_b\nabla_a  k^b = \nabla_a \nabla_b k^b - {  {R_{ab} }^b}_d k^d 
\equiv R_{ad} k^d  \label{caz2o}
\ee
and $\Box k_a= \nabla_b\nabla_ak^b = R_{cd}k^d $.

Thus far, the equations of this section apply whether $\phi$ is a test or 
a gravitating 
field. In the case in which $\phi$ is the only matter source of the 
Einstein equations, one can go further and note that in this special 
situation  $T^{(\phi)}=0$ yields  $R=0$ and the Einstein equations reduce 
to $R_{ab}=8\pi \nabla_a\phi \nabla_b\phi $, from which it follows that 
$R_{ab}k^b = 8\pi k_a k_b k^b=0$ and 
\be
\Box k^c=0 \,.
\ee

\section{Scalar-tensor gravity}
\label{sec:3}
\setcounter{equation}{0}

Let us consider now scalar-tensor gravity. In this case, the 
Brans-Dicke-like field $\Phi$ cannot be a test field and 
always gravitates.  Assuming the gradient $\nabla^c \Phi$ to be null, the 
effective stress-energy tensor $T_{ab}^{(\Phi)}$ cannot be a null dust. In 
fact, assume $V(\Phi) \equiv 0$ and $T^{(m)}=0$, then Eq.~(\ref{BDfe2}) 
gives $\Box\Phi=0$ and~(\ref{Phiset}) reduces to
\be
T_{ab}^{(\Phi)}= \frac{\omega}{\Phi^2} \, \nabla_a \Phi\nabla_b  \Phi 
+\frac{ \nabla_a \nabla_b\Phi}{ \Phi}  \,,
\ee
which satisfies $T^{(\Phi)}=0$. One can set $k_a = \nabla_a \ln \Phi$ and 
\be
T_{ab}^{(\Phi)}=\omega k_a k_b + \frac{ \nabla_a \nabla_b  \Phi}{  \Phi} 
\,.
\ee
Then $k^c $ is null  and is also divergence-free since 
\be
\nabla^c k_c  = \frac{\Box \Phi}{\Phi} -\frac{ \nabla^a \Phi\nabla_a 
\Phi}{ \Phi^2} =0\,,
\ee
where the first term vanishes because of the field equation for $\Phi$. 
Using the identity 
\be
\frac{ \nabla_a \nabla_b \Phi}{ \Phi} = 
 \nabla_{(a} \left( \frac{ \nabla_{b)} \Phi}{ \Phi} \right) 
+ \frac{ \nabla_a \Phi \nabla_b \Phi}{\Phi^2} \,,
\ee
one obtains
\be
T_{ab}^{(\Phi)} = \left( \omega+1\right) k_a k_b + \frac{ \nabla_a k_b + 
\nabla_b k_a}{2} \,,
\ee
which is not the stress-energy tensor of a null dust due to the second 
term on the right hand side. Unless $k^a$ is a (null) Killing vector 
field \cite{Waldbook} satisfying the Killing equation $\nabla_a k_b +
\nabla_b k_a =2\nabla_a \nabla_b \Phi=0$, this second 
term 
does not vanish. It is clear that this restriction 
can be obeyed only by very special geometries and, therefore, we will not 
pursue it further. 

By proceeding in analogy with Sec.~\ref{sec:2}, one derives the wave 
equation for $k^c$
\be
\Box k_a + R_{ab} k^b=0 \label{eq:wv2}
\ee
(note the opposite sign to Eq.~(\ref{eq:wv}) in the coupling to the 
Ricci tensor). In fact, the covariant conservation $ \nabla^b 
T_{ab}^{(\Phi)}=0$ gives
\be
\left( \omega+1 \right) \left( k_a \nabla^b k_b +k^b \nabla_b k_a \right) 
+\frac{1}{2} \left( \Box k_a +\nabla_b \nabla_a k^b \right)=0 \,, 
\ee
where the first bracket vanishes because $k^a$ is divergence-free and 
because of the affinely parameterized geodesic equation. The use of 
Eq.~(\ref{caz2o}) then yields Eq.~(\ref{eq:wv2}).  If, in addition, 
$T^{(m)}=0$, Eq.~(\ref{eq:wv2}) simplifies to $\Box k^c=0$.

\section{Conclusions}
\label{sec:4}
\setcounter{equation}{0}

We have filled a gap in the literature regarding the equivalence between a 
scalar field $\phi$ and an effective null dust when the gradient $\nabla^c 
\phi$ of this scalar is a null vector field over a region of spacetime. A 
canonical, minimally coupled, free and massless scalar field with null 
gradient is 
equivalent to an irrotational null dust. When attempting to generalize 
this property to a gravitating Brans-Dicke-like scalar field $\Phi$ in 
scalar-tensor gravity \cite{BD, ST}, we have found that the equivalence 
does not carry over, unless the null gradient of $\Phi$ is also a Killing 
vector. This is a very strong restriction, which makes this situation 
rather uninteresting from the physical point of view.

\begin{acknowledgments}

This work is supported, in part, by the Natural Sciences and Engineering 
Research Council of Canada (Grant No.~2016-03803 to V.F.).

\end{acknowledgments}


\end{document}